\documentclass[twocolumn]{revtex4}
\usepackage{epsfig}
\usepackage{graphicx}
\usepackage{amssymb}
\usepackage{amsmath}
\usepackage{amsthm}
\usepackage{amsfonts}
\usepackage{mathrsfs}
\usepackage[dvips]{color}
\usepackage{multirow}

\newcommand{\fn}{{\,\mathfrak{n}\,}}

\newcommand{\bI}{\mathbf{I}}

\newcommand{\bM}{\mathbf{M}}

\newcommand{\cP}{\mathcal{P}}

\newcommand{\cR}{\mathcal{R}}

\newcommand{\cT}{\mathcal{T}}

\newcommand{\be}{\begin{equation}}
\newcommand{\ee}{\end{equation}}
\newcommand{\bea}{\begin{eqnarray}}
\newcommand{\eea}{\end{eqnarray}}
\newcommand{\nn}{\nonumber}

\newcommand{\ed}{\end{document}}

\newcommand{\bi}{\begin{itemize}}
\newcommand{\ei}{\end{itemize}}

\newcommand{\bce}{\begin{center}}
\newcommand{\ece}{\end{center}}

\newcommand{\sR}{\mathscr{R}}

\newcommand{\sT}{\mathscr{T}}

\begin{document}

\title{Unidirectionally Invisible Potentials as Local Building Blocks of\\ all Scattering Potentials}

\author{Ali~Mostafazadeh}
\address{Departments of Physics and Mathematics, Ko\c{c} University, Sar{\i}yer 34450, Istanbul, Turkey\\
amostafazadeh@ku.edu.tr}

\begin{abstract}
We give a complete solution of the problem of constructing a scattering potential $v(x)$ that possesses scattering properties of one's choice at an arbitrary prescribed wavenumber. Our solution involves expressing $v(x)$ as the sum of at most six unidirectionally invisible finite-range potentials for which we give explicit formulas. Our results can be employed for designing optical potentials. We discuss its application in modeling threshold lasers, coherent perfect absorbers, and bidirectionally and unidirectionally reflectionless absorbers, amplifiers, and phase shifters.



\end{abstract}

\maketitle

Complex scattering potentials in one dimension provide a fertile ground for modeling various active optical systems. Among these are systems displaying spectral singularities \cite{prl-2009,ss1,ss2,ss3} and unidirectional reflectionlessness and invisibility \cite{lin,invisible,feng-yin,pra-2013a,pra-2014}. Spectral singularities correspond to scattering states that behave exactly like zero-width resonances \cite{prl-2009}. In optics they give rise to lasing at the threshold gain \cite{pra-2011a} while their time-reversal is responsible for coherent perfect absorption (CPA) or antilasing \cite{antilasing,antilasing2}. Unidirectional reflectionlessness and invisibility are also of great interest, because they offer means of realizing one-way linear optical devices \cite{lin,feng-yin}. The task of designing scattering potentials that support such desirable properties is clearly a problem of basic theoretical and practical importance. In this article we give a complete solution for this problem.

Throughout this investigation, we use the transfer matrix of one-dimensional scattering theory \cite{sanchez} as our main tool. For a scattering potential $v(x)$, the general solution of the Schr\"odiger equation, $-\psi''(x)+v(x)\psi(x)=k^2\psi(x)$, or the Helmholtz equation, $\psi''(x)+k^2\fn(x)^2\psi(x)=0$, has the asymptotic form:
	\be
	\psi_{\pm}(x)=A_\pm e^{ikx}+B_\pm e^{-ikx}~~{\rm for}~~x\to\pm\infty,
	\label{eq01}
	\ee
where $k$ is the wavenumber, $\fn(x)$ is a refractive index that we can relate to $v(x)$ via $\fn(x)=\sqrt{1-v(x)/k^2}$, and $A_\pm$ and $B_\pm$ are complex coefficients. The transfer matrix of $v(x)$ (and $\fn(x)$) is the $2\times 2$ matrix $\bM$ satisfying
    \be
    \bM\left[\begin{array}{c} A_-\\ B_-\end{array}\right]=\left[\begin{array}{c} A_+\\ B_+\end{array}\right].
    \label{def-M}
    \ee
We can express it in terms of the left and right reflection amplitudes, $R^l$ and $R^r$, and the transmission amplitude $T$ of $v(x)$ according to \cite{prl-2009}:
    \be
    \bM=\left[\begin{array}{cc}
    T- R^lR^r/T & R^r/T\\
    - R^l/T & 1/T\end{array}\right].
    \label{M=}
    \ee
Recall that the asymptotic scattering solutions of the above Schr\"odinger and Helmholtz equations have the form
    \bea
    \psi_l(x)&=&\left\{\begin{array}{ccc}
    e^{ikx}+R^l\,e^{-ikx}&{\rm for}& x\to-\infty,\\
    T\,e^{ikx}&{\rm for}& x\to+\infty,\end{array}\right.\nn\\
    \psi_r(x)&=&\left\{\begin{array}{ccc}
    T\,e^{-ikx}&{\rm for}& x\to-\infty,\\
    e^{-ikx}+R^r\,e^{ikx}&{\rm for}& x\to+\infty,\end{array}\right.\nn
    \eea
and the reflection and transmission coefficients are given by $|R^{l/r}|^2$ and $|T|^2$.

We can use (\ref{eq01}), (\ref{def-M}) and (\ref{M=}) to derive the following transformation rules for $R^{l/r}$ and $T$ under the space reflection (parity), $x\stackrel{\cP}{\rightarrow}-x$, space translations, $x\stackrel{T_a}{\longrightarrow}x-a$, and the time-reversal transformation, $\psi(x)\stackrel{\cT}{\rightarrow}\psi(x)^*$.
    \begin{align}
    &R^{l}\stackrel{\cP}{\longrightarrow} R^{r},
    && R^{r}\stackrel{\cP}{\longrightarrow} R^{l},
    && T\stackrel{\cP}{\longrightarrow} T,
    \label{R-T-trans1}\\
    &R^{l}\stackrel{T_a}{\longrightarrow} e^{2i ak} R^{l}, &&
    R^{r}\stackrel{T_a}{\longrightarrow} e^{-2i ak} R^{r},
    &&T\stackrel{T_a}{\longrightarrow} T,
    \label{R-T-trans3}\\
    &R^{l}\stackrel{\cT}{\longrightarrow}-\frac{R^{r*}}{D^*},
    &&R^{r}\stackrel{\cT}{\longrightarrow}-\frac{R^{l*}}{D^*},
    && T  \stackrel{\cT}{\longrightarrow} \frac{T^{*}}{D^*}.
    \label{R-T-trans2}
    \end{align}
    where $D:=T^2-R^lR^r$.

We say that $v(x)$ is unidirectionally reflectionless at a wavenumber $k_0$, if for $k=k_0$, $R^{l}=0\neq R^{r}$ or $R^{r}=0\neq R^{l}$. We respectively refer to these conditions as ``left-reflectionlessness'' and ``right-reflectionlessness.'' Clearly the condition for bidirectional reflectionlessness is $R^l=R^r=0$. If a potential that is unidirectionally reflectionless at a wavenumber $k_0$ has a unit transmission amplitude, i.e., $T=1$, at this wavenumber, we call it ``unidirectionally invisible'' \cite{lin}. Similarly, we use ``left-invisible'', ``right-invisible'', and ``bidirectionally invisible'' to mean that a potential is respectively left-reflectionless, right-reflectionless, and bidirectionally reflectionless and in addition satisfies $T=1$ at $k=k_0$.

In this article we use the properties of the transfer matrix to address the problem of constructing scattering potentials with specific scattering properties at a given real and positive wavenumber. For future reference we use the term ``local inverse scattering'' to refer to this problem. We also make use of the notion of the support of a potential $v(x)$ which is the smallest closed interval ${I}$ outside which $v(x)$ vanishes. For a pair of potentials, $v_1(x)$ and $v_2(x)$, with support ${I}_1$ and ${I}_2$, we use the notation ``${I}_1\prec{I}_2$'' to means that ${I}_1$ lies to the left of ${I}_2$, i.e., for all $x_1\in{I}_1$ and $x_2\in{I}_2$, we have $x_1\leq x_2$. Suppose that  ${I}_1\prec{I}_2$ and $v(x)=v_1(x)+v_2(x)$. Then we can use (\ref{def-M}) to relate $\bM$ to the transfer matrices $\bM_i$ of $v_i(x)$ according to $\bM=\bM_2\bM_1$. This, so-called composition property, makes the transfer matrix an extremely useful tool in dealing with a variety of physics and engineering problems \cite{sanchez}.

If $v(x)$ is a finite-range potential, ${I}=[a,b]$, and we can write $v(x)=g(x)\theta(x-a)\theta(b-x)$, where $g(x)$ gives the value of the potential for $x\in[a,b]$, and $\theta(x)$ is the Heaviside step function; $\theta(x):=0$ for $x<0$, and $\theta(x):=1$ for $x\geq 0$.

We begin our analysis by considering the local inverse scattering problem for potentials with finite reflection and transmission amplitudes, $R^{l/r}$ and $T$, at the prescribed wavenumber $k_0$. Suppose that we can construct finite-range potentials $v_\pm(x)$ and $v_0(x)$ with support $I_\pm$  and $I_0$ and transfer matrices $\bM_\pm$ and $\bM_0$, such that $ I_-\prec I_0\prec I_+$ and
     \bea
     &\bM_-=
     \left[\begin{array}{cc}
     1 & 0 \\
     -R^l  & 1\end{array}\right],
     ~~~~~\bM_+=\left[\begin{array}{cc}
     1& R^r \\
     0 & 1\end{array}\right],&
     \label{M=pm}\\
     &\bM_0=\left[\begin{array}{cc}
     T& 0 \\
     0 & 1/T\end{array}\right],&
     \eea
for $k=k_0$. Then the transfer matrix of the potential
    \be
    v(x):= v_-(x)+v_0(x)+v_+(x)
    \label{potential}
    \ee
is given by
            $\bM_+ \bM_0\, \bM_-=\left[\begin{array}{cc}
            T- R^lR^r/T & R^r/T\\
            - R^l/T & 1/T\end{array}\right]$.
Comparing this relation with (\ref{M=}), we see that $R^l$, $R^r$, and $T$ are respectively the left reflection, right reflection, and transmission amplitudes of $v(x)$ at the wavenumber $k_0$. This reduces the local inverse scattering problem for potentials with finite reflection and transmission amplitudes to the construction of the bidirectionally reflectionless potential $v_0(x)$ and the unidirectionally invisible potentials $v_\pm(x)$.

Next, we show that $v_0(x)$ can be constructed out of four unidirectionally invisible potentials.

Let $\cR$ be an arbitrary nonzero complex number, and for each $j\in\{1,2,3,4\}$, $v_j(x)$ be a finite-range unidirectionally invisible potential with support ${I}_j$ and reflection amplitudes $R_j^{l/r}$ at wavenumber $k_0$ such that $I_1\prec I_2\prec I_3 \prec I_4$, $R^r_1=R^l_2=R^r_3=R^l_4=0$, and
	\begin{align*}
	& R^l_1=-\cR T, && R^r_2=\frac{T-1}{\cR T}, && R^l_3=\cR, && R^r_4=\frac{1-T}{\cR}.
	\end{align*}
Then $v_1(x)$ and $v_3(x)$ are right-invisible while $v_2(x)$ and $v_4(x)$ are left-invisible, and the transfer matrix $\bM_j$ of $v_j(x)$ at $k_0$ has the form
	\begin{align}
	&\bM_1=\left[\begin{array}{cc}
	1 & 0\\[9pt]
	\cR\, T & 1\end{array}\right],
    && \bM_2=\left[\begin{array}{cc}
	1 & (T-1)/\cR T\\
	0 & 1\end{array}\right],
    \label{q201}\\
    & \bM_3=\left[\begin{array}{cc}
	1 &  0\\
	-\cR & 1\end{array}\right],
    &&\bM_4=\left[\begin{array}{cc}
	1 & (1-T)/\cR\\
	0 & 1\end{array}\right].
    \label{q202}
	\end{align}
It is an easy exercise to show that for $k=k_0$ the transfer matrix of the potential $\sum_{j=1}^4 v_j(x)$ is given by
	$\bM_4 \bM_3 \bM_2 \bM_1=\left[\begin{array}{cc}
	T & 0 \\
	0 & 1/T\end{array}\right]$.
This shows that we can take
    \be
    v_0(x)=v_1(x)+v_2(x)+v_3(x)+v_4(x),
    \label{v0=}
    \ee
i.e., we can construct a bidirectionally reflectionless potential by assembling four unidirectionally invisible potentials.

Next, we examine the problem of constructing a scattering potential whose reflection and transmission amplitudes diverge at a prescribed real and positive wavenumber $k_0$.

Let  $w_\pm(x)$ be a pair of scattering potentials with support $J_\pm$ such that for the wavenumber $k_0$, $w_\pm(x)$ has the same transfer matrix as $v_\pm(x)$, but $J_+\prec J_-$. Then, in view of (\ref{M=pm}), the transfer matrix of the potential
	\be
	w(x):=w_-(x)+w_+(x),
	\label{w=}
	\ee
at $k_0$ is given by $\bM_-\bM_+=\left[\begin{array}{cc}
      1 & \displaystyle R^r\\
       -R^l  & 1-R^lR^r\end{array}\right]$.
Comparing this relation with (\ref{M=}), we see that the transmission amplitude of $w(x)$ diverges whenever
    \be
    R^l R^r=1.
    \label{R-R}
    \ee
If this condition holds, the reflection amplitudes of $w(x)$ also diverge for $k=k_0$, $k_0^2$ is a spectral singularity \cite{prl-2009} of the potential $w(x)$, and $w(x)^*$ displays CPA at $k_0$, \cite{antilasing2}.

The arguments we have so far presented show that we can realize any kind of scattering effect at a given wavenumber using certain finite-range potentials that are the sum of up to six unidirectionally invisible potentials. This solves the local inverse scattering problem for general potentials provided that we can construct the constituent unidirectionally invisible potentials, namely $v_\pm(x)$, $v_j(x)$, and $w_\pm(x)$. In the following we achieve this by introducing a particular family of finite-range unidirectionally invisible potentials.

First, we examine the case of right-invisible potentials, i.e., for a given nonzero complex number $R$, we construct a scattering potential whose right reflection, left reflection, and transmission amplitudes at a prescribed wavenumber $k_0$ are respectively given by $R^r=0$, $R^l=R$, and $T=1$.

Consider the finite-range potentials of the form,
    \bea
    &&v_{\alpha,n}(x):=k^2 f_{\alpha}(x)\theta(x)\theta(L_n-x),
    \label{q11}\\
    &&f_\alpha(x):=\frac{-8\,\alpha (3-2 e^{2ik_0x})}{e^{4ik_0x}+\alpha(1-e^{2ik_0x})^2}.
    \label{f=}
	\eea
where $\alpha$ and $n$ take real and positive integer values, respectively, and $L_n:=\pi n/k_0$.
As we show in the appendix, $v_{\alpha,n}(x)$ is right-invisible at $k=k_0$, and for $\alpha>-1/4$ its left reflection amplitude at $k_0$ has the form
	\be
	R^l=R^l_{\alpha,n}:=\frac{-8 \pi i n\, \alpha}{(\alpha+1)^3}.
	\label{q13-n}
	\ee

Because $|\alpha|$ and $n$ can respectively take arbitrarily small and large positive values, we can always choose them such that $8\pi n\alpha/(\alpha+1)^3=|R|$. Let $\varphi\in[0,2\pi)$ denote the phase angle of $R$, so that $R=|R| e^{i\varphi}$. Then (\ref{R-T-trans3}) implies that for every integer $m$, the translation $x\to x+d$ with
    \be
    d:=\frac{(4m-1)\pi-2\varphi}{4k_0},
    \label{d=123}
    \ee
maps $v_{\alpha,n}(x)$ to the potential,
    \be
    v^r_{R}(x):=v_{\alpha,n}(x+d),
    \label{v-m-R}
    \ee
that is right-invisible at $k=k_0$, and its left reflection amplitude at this wavenumber coincides with $R$. This completes the solution of the local inverse scattering problem for right-invisible potentials. Because every left-invisible potential can be written as the complex-conjugate of a right-invisible potential (see (\ref{R-T-trans2})), the above construction solves the local inverse scattering problem for all unidirectionally invisible potentials.

$v^r_{R}(x)$ and $v^r_{R}(x)^*$ are finite-range potentials with support $[-d,L_n-d]$. Because $m$ is an arbitrary integer, we can choose it so that the support of these potentials lie to the left or right of any given finite interval. This shows that making appropriate choices for $\alpha, n,\varphi$, and $m$, we can use  $v^r_{R}(x)$ and $v^r_{R}(x)^*$ to obtain an explicit realization of the potentials $v_\pm(x)$, $v_j(x)$, and $w_\pm(x)$ that appear in (\ref{potential}), (\ref{v0=}), and (\ref{w=}). This completes our solution of the local inverse scattering problem for general scattering potentials.

As a concrete example, let us identify the potential $w_+(x)$ with $v_{\alpha,n}(x+L_n)^*$, where $v_{\alpha,n}(x)$ is given by (\ref{q11}), $\alpha>-1/4$, and $n$ is a positive integer. Then the support of $w_+(x)$ is $[-L_n,0]$. According to (\ref{R-T-trans3}),  $w_+(x)$ has the same transmission amplitude as $v_{\alpha,n}(x)^*$, and that its left (respectively right) reflection amplitude differs from that of $v_{\alpha,n}(x)^*$ by a phase factor $e^{-2i kL_n}$ (respectively  $e^{2i kL_n}$). But because $kL_n$ is an integer multiple of $\pi$, $e^{\pm2i kL_n}=1$. This together with (\ref{R-T-trans2}) and (\ref{q13-n}) implies that, at  $k=k_0$, $T=1$, $R^l=0$, and
    \be
    R^r=-R^{l*}_{\alpha,n}=R^{l}_{\alpha,n}=
    \frac{-8 \pi i n\, \alpha}{(\alpha+1)^3}.
    \label{eq17new}
    \ee

Next, we identify $w_-(x)$ with $v_{\beta,m}(x)$ where $\beta>-1/4$ and $m$ is a positive integer. Then $w_-(x)$ is a potential with support $[0,L_m]$ that is right-invisible at $k=k_0$, and in light of (\ref{q13-n}) its left reflection amplitude at this wavenumber is given by
    \be
    R^l=R^l_{\beta,m}=\frac{-8 \pi i \beta m}{(\beta+1)^3}.
    \label{eq17new2}
    \ee
Substituting this relation and (\ref{eq17new}) in (\ref{R-R}) gives
    \be
    (\alpha+1)^3(\beta+1)^3+64\pi^2 m n\alpha\beta=0.
    \label{q14}
    \ee
It turns out that for each nonzero $\alpha$ greater than $-1/4$ and positive integers $m$ and $n$, this equation can be solved to find values for $\beta$ that is greater than $-1/4$. In particular, for situations where $|\alpha|\ll 1$, there is a single value of $\beta$ satisfying $|\beta|\ll 1$. For this value of $\beta$, the above construction works and the potential (\ref{w=}) has a spectral singularity at the wavenumber $k_0$. In view of (\ref{q11}) and (\ref{f=}) the refractive index $\fn(x)$ associated with this potential satisfies
    \be
    \fn^{\! 2}(x)=\left\{\begin{array}{cc}
    \!\!1+\displaystyle \frac{8\,\alpha (3-2 e^{-2ik_0x})}{e^{-4ik_0x}+\alpha(1-e^{-2ik_0x})^2} & \!\!{\rm for}~x\in[-L_n,0),\\[9pt]
    \!\! 1+\displaystyle \frac{8\,\beta (3-2 e^{2ik_0x})}{e^{4ik_0x}+\beta(1-e^{2ik_0x})^2} & {\rm for}~x\in[0,L_m],\\[9pt]
    1 & {\rm otherwise}.\end{array}\right.
    \label{fg1}
    \ee
An optical system given by the complex-conjugate of this refractive index profile displays CPA at $k=k_0$.

\begin{figure}
	\begin{center}
	\includegraphics[scale=.6]{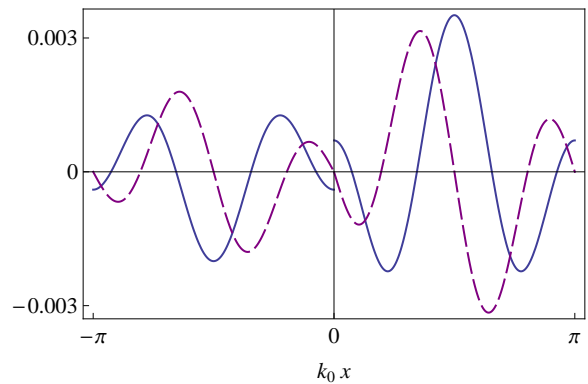}
	\caption{(Color online) Graphs of real part of of $\fn(x)-1$ (solid blue curve) and the imaginary part of $\fn(x)$ (dashed purple curve) for $\fn(x)$ given by (\ref{fg1}) with $\alpha=-10^{-4}$, $\beta=1.759\times 10^{-4}$, and $|k_0x|\leq \pi$.}
	\label{fig1}
	\end{center}
	\end{figure}

\begin{figure}
	\begin{center}
	\includegraphics[scale=.65]{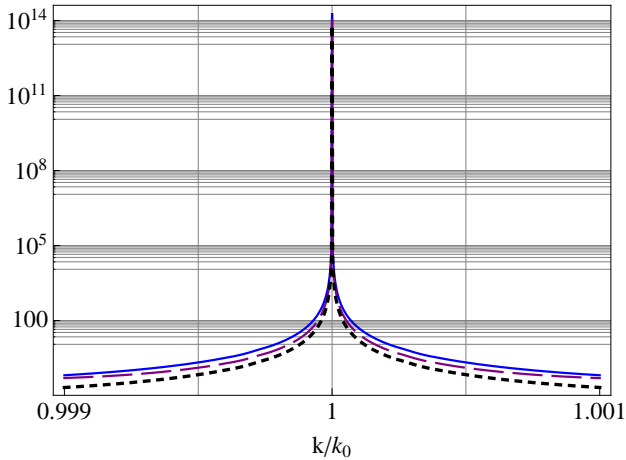}
	\caption{(Color online) Graphs of the reflection and transmission coefficients, $|R^{l}|^2$ (blue full curve), $|R^r|^2$ (black dotted curve) and $|T|^2$ (purple dashed curve), as functions of $k/k_0$ for the refractive index profile given by (\ref{fg1}) with $\alpha=-10^{-4}$ and $\beta=1.759\times 10^{-4}$. The sharp peak at $k=k_0$ is an obvious signature of a spectral singularity.}
	\label{fig2}
	\end{center}
	\end{figure}

For example, let us take $\alpha=-10^{-4}$ and $m=n=300$. Then (\ref{q14}) gives $\beta=1.759\times 10^{-4}$. These numerical values are consistent with the fact that in typical optical applications $|\fn_{\alpha,n}-1|$ and $|\fn_{\beta,m}-1|$ are at most of the order of $10^{-3}$. Fig.~\ref{fig1} shows the plots of real and imaginary parts of $\fn(x)-1$ for these choices of $\alpha$ and $\beta$, and $|k_0x|\leq \pi$.
If we take $k_0=2\pi~(\mu{\rm m})^{-1}$ (wavelength=$1\,\mu{\rm m}$), $L_m=L_n=300\pi/k_0=150\mu{\rm m}$. Therefore, this setup corresponds to an infinite planar slab of inhomogeneous optically active material of thickness $300~\mu{\rm m}$. Substituting these numerical values for $\alpha,\beta,m$, and $n$ in (\ref{eq17new}) and (\ref{eq17new2}), we find $R^l=0.754 i$ and $R^r=-1.323 i$. These match the numerical calculation of these quantities. Fig.~\ref{fig2} shows the logarithmic plots of the reflection and transmission coefficients for this model. The extremely sharp peak at $k=k_0$ is a clear evidence of the presence of a spectral singularity.

The ability to use finite-range unidirectionally invisible potentials to realize arbitrary scattering effects provides a powerful method for designing optical potentials. For example, consider the construction of the bidirectionally reflectionless potential $v_0(x)$ in terms of the unidirectionally invisible potentials $v_j(x)$. Suppose that we take
    \begin{align}
    &R^l_1=-\epsilon, &&R^r_2=e^{i\theta} \epsilon-\epsilon^{-1},
    \label{q600}\\
    &R^l_3=e^{-i\theta} \epsilon^{-1}, && R^r_4=e^{i\theta}\epsilon-e^{2i\theta}\epsilon^{3},
    \label{q601}
    \end{align}
where $\epsilon$ is a positive real number and $\theta\in[0,2\pi)$. Then the potential (\ref{v0=}) is a bidirectionally reflectionless potential with transmission amplitude $T=\epsilon^2 e^{i\theta}$. We can use the potential (\ref{v-m-R}) and its complex-conjugate to realize $v_j(x)$ for arbitrary values of $\epsilon$. In particular, for $\epsilon<1$, $v_0(x)$ models a bidirectionally reflectionless absorber that operates at the wavenumber $k_0$ and induces a specific phase shift in the transmitted wave. Notice that for $\epsilon\ll 1$, $|R^r_4|\approx|R^l_1|=\epsilon$ and $|R^r_2|\approx|R^l_3|=\epsilon^{-1}$. Therefore, we obtain a quadratic absorption effect corresponding to $|T|=\epsilon^2$ using unidirectionally invisible amplifiers, $v_1(x)$ and $v_4(x)$, and unidirectionally invisible absorbers, $v_2(x)$ and $v_3(x)$, whose strength is linear in $\epsilon$.

Clearly, for $\epsilon=1$, the above system operators as a bidirectionally reflectionless phase shifter, and for $\epsilon>1$ it serves as a bidirectionally reflectionless transmission amplifier that produces a specific phase shift. We can also construct unidirectionally reflectionless absorbers, amplifiers, and phase-shifters. This follows from the observation that we can construct unidirectionally reflectionless potentials by adding to $v_0(x)$ a finite-range unidirectionally invisible potential $u(x)$ whose support $J$ lies to the left or right of $I_0$. Specifically, if $u(x)$ is left- (respectively right-) invisible and $I_0\prec J$ (respectively $J\prec I_0$), $v_0(x)+u(x)$ is left- (respectively right-) reflectionless.

\begin{figure}
	\begin{center}
	\includegraphics[scale=.68]{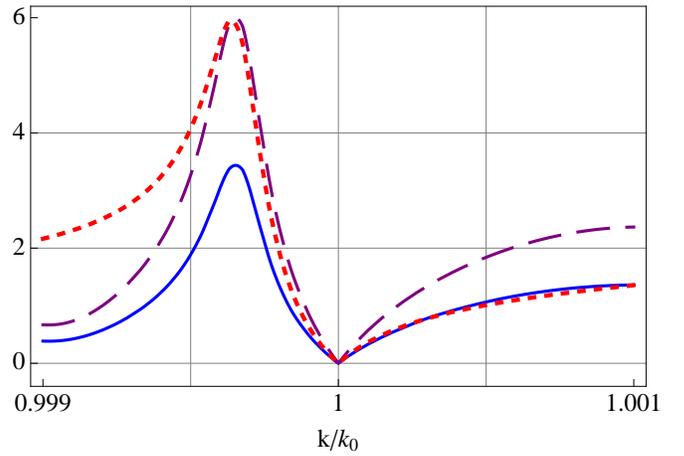}
	\caption{(Color online) Graphs of $|R^{l}|$ (solid blue curve) and $|R^r|$ (dashed purple curve), and $|T-\sqrt 2 i|$ (dotted red curve) as functions of $k/k_0$ for the potential given by (\ref{v0=}) and (\ref{v0=sp}), and $k_0=2\pi/\mu {\rm m}$. The fact that all of these quantities vanish for $k=k_0$ shows that this potential models a bidirectionally reflectionless amplifier that doubles the intensity of the incident wave upon transmission and induces a $\pi/2$-phase.}
	\label{fig3}
	\end{center}
	\end{figure}

For example suppose that we wish use the above construction to obtain a bidirectionally reflectionless amplifier that doubles the intensity of the incident wave ($|T|^2=2$) and induces a $\pi/2$-phase shift at the wavelength $1~\mu{\rm m}$ ($k_0=2\pi/\mu{\rm m}$) upon transmission. This corresponds to setting $\epsilon=2^{1/4}$ and $\theta=\pi/2$ in (\ref{q600}) and (\ref{q601}). A potential $v_0(x)$ that achieves this goal is given by (\ref{v0=}) where
    \be
    v_j(x):=\left\{\begin{array}{ccc}
    v_{\alpha_j,n}(x+d_j) &{\rm for} &j=1,3,\\
    v_{\alpha_j,n}(x+d_j)^* &{\rm for} &j=2,4,
    \end{array}\right.
    \label{v0=sp}
    \ee
$n=300$, so that $L_n=150~\mu{\rm m}$, and
    \begin{align*}
    &\alpha_1=1.57798\times 10^{-4}, && d_1=300.625~\mu{\rm m},\\
    &\alpha_2=1.93283\times 10^{-4}, && d_2=150.299~\mu{\rm m},\\
    &\alpha_3=1.11565\times 10^{-4}, && d_3=0.00000~\mu{\rm m},\\
    &\alpha_4=2.73409\times 10^{-4}, && d_4=-150.326~\mu{\rm m}.
    \end{align*}
In particular, the support $I_j$ of $v_j(x)$ that is given by $[-d_j,L_n-d_j]$ fulfils the requirement $I_1\prec I_2\prec I_3 \prec I_4$. $v_0(x)$ corresponds to a particular gain-loss profile that is confined in a planar slab of thickness $600.951\mu{\rm m}$. Figure~\ref{fig3} shows the plots of $|R^l|$, $|R^r|$ and $|T-\sqrt 2 i|$ for this system. The vanishing of all these quantities at $k=k_0$ is a graphical (numerical) confirmation of our analytical results.

In conclusion, we have revealed an intriguing property of unidirectionally invisible potentials, namely that using at most six such potentials we can realize any scattering effects at any wavenumber. This solution allows for designing a variety of optical potentials that can model threshold lasers, antilasers, and bidirectionally and unidirectionally reflectionless absorbers, amplifiers, and phase shifters. The main disadvantage of our construction is that it yields potentials that realize the desired scattering properties only at a single wavenumber. However, it also implies that these properties are essentially preserved in the spectral bands in which the constituting unidirectionally invisible retain their scattering features. This reduces the general inverse scattering problem for arbitrary real or complex scattering potentials to constructing broadband unidirectionally invisible potentials. The study of this kind of potentials can be conducted using the standard inverse scattering theory \cite{IS} which is typically difficult to apply, for it involves solving certain integral equations. In this context, it might be useful to note the possibility of constructing broadband unidirectionally reflectionless bilayer slabs that operate in spectral bands as wide as 100~nm \cite{pra-2013a}. Finally, we wish to note that our approach allows us to construct potentials that display a particular scattering effect at more than one wavenumber. This requires using constituent unidirectionally invisible potentials that have the appropriate properties at these wavenumbers. The problem of constructing finite-range potentials displaying perturbative unidirectional invisibility at finitely or infinitely many wavenumbers is treated in Ref.~\cite{pra-2014}.

\vspace{6pt}

\noindent{\em Acknowledgments:}  This work has been supported by  the Scientific and Technological Research Council of Turkey (T\"UB\.{I}TAK) in the framework of the project no: 112T951, and by the Turkish Academy of Sciences (T\"UBA).

\vspace{6pt}

\noindent{\em Appendix:} In Ref.~\cite{ap-2014} we show that the reflection and transmission amplitudes of a finite-range potential $v(x)$ with support $[0,L]$ are given by
    \bea
    R^l&=&-\int_{1}^{z_+} dz~\frac{S''(z)}{S(z)S'(z)^2},
	\label{e2nn}\\
	R^r&=&\frac{S(z_+)}{S'(z_+)}-z_+,~~~T=\frac{1}{S'(z_+)},
	\label{e3nn}
	\eea
where $z:=e^{-2ik x}$, $z_+:=e^{-2ik L}$, $S(z)$ is the solution of the initial-value problem:
	\bea
	z^2S''(z)+\left[\frac{\check v(z)}{4k^2}\right]S(z)=0,
    ~~~S(1)=S'(1)=1,~~~
    \label{e1n}
	\eea
$\check v(z):=v(i\ln z/(2k))=v(x)$, and the integral in (\ref{e2nn}) is to be evaluated along the clockwise-oriented circular arc: $\{e^{-2i t}|t\in[0,kL]\}$. In particular, whenever $kL=\pi n$ for some positive integer $n$, $z_+=1$, and the integral in (\ref{e2nn}) becomes a contour integral that is to be evaluated along the clockwise-oriented contour $\gamma_n:=\{ e^{-i t}~|~t\in[0,2\pi n]\}$, i.e., the $n$-fold covering of the unit circle $\gamma_1$. In this case, $R^r=0$, $T=1$, and $v(x)$ is a right-invisible potential provided that $R^l\neq 0$. The latter relation holds whenever the integrand in (\ref{e2nn}) has poles inside $\gamma_1$, and the sum of the residues of the integrand at these poles is nonzero.

The potential $v_{\alpha,n}(x)$ introduced in (\ref{q11}) corresponds to setting $k=k_0$, $L=n\pi/k_0$, and $S(z)=z[\alpha(z-1)^2+1]$. This choice for $S(z)$ fulfils (\ref{e1n}) if we set $v(x)=v_{\alpha,n}(x)$. It is also easy to check that this potential is smooth on its support except for $\alpha=-1/4$ where it develops a singularity at $x=(2\ell+1)\pi/2k$ for integer $\ell$. Furthermore, for $\alpha>-1/4$, the integrand in (\ref{e2nn}) has a single pole lying inside $\gamma_1$ (namely $z=0$). In this case, we can easily use the Residue Theorem to evaluate the right-hand side of (\ref{e2nn}). This results in (\ref{q13-n}).

\section*{Addendum}

This addendum is prepared after the above paper was published as [A.~Mostafazadeh, Phys.\ Rev.\ A~{\bf 90}, 023833 (2014)]. In the following we refer it as [$\star$]. The addendum is to appear as a separate brief report in Phys.\ Rev.\ A. It improves the main result of [$\star$] by reducing the maximum number $n$ of the finite-range unidirectionally invisible potentials necessary for constructing a scattering potential $v(x)$ with given scattering properties at a prescribed wavenumber from $6$ to $4$. In particular, we show that we can construct $v(x)$ as the sum of up to $n=3$ finite-range unidirectionally invisible potentials, unless if it is required to be bidirectionally reflectionless.\vspace{12pt}

In Ref.~[$\star$], we employ the composition property of the transfer matrix of one-dimensional potential scattering theory to give a complete solution of a local inverse scattering problem that aims at constructing a scattering potential $v(x)$ with given (left and right) reflection and transmission amplitudes, $R^{l/r}$ and $T$, at a prescribed wave number $k_0$. This solution involves expressing $v(x)$ in the form
    \be
    v(x)=v_-(x)+v_1(x)+v_2(x)+v_3(x)+v_4(x)+v_+(x),
    \label{v=sum}
    \ee
where $v_\pm(x)$ and $v_j(x)$ (with $j=1,2,3,4$) either vanish or are unidirectionally invisible finite-range potentials with support $I_\pm$ and $I_j$, and transfer matrix  $\bM_\pm$ and $\bM_j$, such that
   \be
   I_-\prec I_1\prec I_2\prec I_3\prec I_4\prec I_4,
   \label{Is}
   \ee
($I\prec J$ means that $I$ lies on the left of $J$, i.e., for all $x\in I$ and $y\in J$, $x\leq y$), for $k=k_0$,
    \begin{align}
    &\bM_-=
     \left[\begin{array}{cc}
     1 & 0 \\
     -R^l  & 1\end{array}\right],
    &&\bM_+=\left[\begin{array}{cc}
     1& R^r \\
     0 & 1\end{array}\right],&
     \label{q200}\\
	&\bM_1=\left[\begin{array}{cc}
	1 & 0\\[9pt]
	\cR\, T & 1\end{array}\right],
    && \bM_2=\left[\begin{array}{cc}
	1 & (T-1)/\cR T\\
	0 & 1\end{array}\right],
    \label{q201}\\
    & \bM_3=\left[\begin{array}{cc}
	1 &  0\\
	-\cR & 1\end{array}\right],
    &&\bM_4=\left[\begin{array}{cc}
	1 & (1-T)/\cR\\
	0 & 1\end{array}\right],
    \label{q202}
	\end{align}
and $\cR$ is an arbitrary nonzero complex number. The claim that, at $k=k_0$, the reflection and transmission amplitudes of the potential (\ref{v=sum}) respectively coincide with $R^{l/r}$ and $T$ follows from the fact that the transfer matrix of this potential is give by
    \be
    \bM=\bM_+\bM_4\bM_3\bM_2\bM_1\bM_-=\left[\begin{array}{cc}
    T-R^lR^r/T & R^r/T\\[6pt]
    -R^l/T & 1/T\end{array}\right].
    \label{M=1}
    \ee

Let $\tilde v_1(x)$ and $\tilde v_4(x)$ be a pair of finite-range potentials with support $\tilde I_{1}$ and $\tilde I_{4}$, and transfer matrix $\tilde\bM_{1}$ and $\tilde\bM_{4}$ such that $\tilde I_-\prec I_2$, $I_3\prec\tilde I_+$, and
    \begin{align}
    &\tilde \bM_1=\bM_1\bM_-=\left[\begin{array}{cc}
	1 & 0\\
	-R^l+\cR T & 1\end{array}\right]\\[3pt]
    & \tilde \bM_4=\bM_+\bM_4=\left[\begin{array}{cc}
	1 & \displaystyle R^r+\frac{1-T}{\cR}\\[3pt]
	0 & 1\end{array}\right].
    \end{align}
Then $\tilde v_1(x)$ and $\tilde v_4(x)$ are respectively right- and left-invisible potentials, and  according to (\ref{M=1}), we have $\bM=\tilde\bM_4\bM_3\bM_2\tilde\bM_1$. In light of this relation, for $k=k_0$, the (left and right) reflection and transmission amplitudes of the one-parameter family of potentials:
    \be
    v_{\cR}(x):=\tilde v_1(x)+v_2(x)+v_3(x)+\tilde v_4(x),
    \label{v=sum-4}
    \ee
coincide with $R^{l/r}$ and $T$, respectively.

If $R^l=R^r=0$, i.e., $v(x)$ is bidirectionally reflectionless, we have $\bM_\pm=\bI$, where $\bI$ stands for the $2\times 2$ identity matrix. This shows that we can take $\tilde v_1=v_1$ and $\tilde v_4=v_4$.

If $R^l\neq 0$, we can set $\cR=R^l/T$, so that $\tilde\bM_1=\bI$. This suggests taking $\tilde v_1=0$ which in turn implies $v_{\cR}(x)=v_2(x)+v_3(x)+\tilde v_4(x)$ with $\tilde v_4$ being a left-reflectionless potential with the right reflection amplitude $R^r+T(1-T)/R^l$ at $k=k_0$. Similarly, for $R^r\neq 0$, the choice $\cR=(T-1)/R^r$ allows us to take $\tilde v_4=0$ and $v_{\cR}(x)=\tilde v_1(x)+v_2(x)+v_3(x)$, where $\tilde v_1$ is a right-reflectionless potential with the left reflection amplitude at $k=k_0$ given by $R^l+T(1-T)/R^r$.

These observations together with the fact that there is an explicitly constructible model for arbitrary unidirectionally invisible finite-range potentials [$\star$] prove the following result.\vspace{6pt}\\
    \noindent {\em Theorem:} Let $k_0$ be a positive real number, and $\sR^l$, $\sR^r$, and $\sT$ be arbitrary complex numbers such that $\sT\neq 0$. Then there is a finite-range potential $v (x)$ with the following properties.
        \begin{enumerate}
        \item The reflection and transmission amplitudes of $v(x)$ at $k=k_0$ are respectively given by $R^{l/r}=\sR^{l,r}$ and $T=\sT$;
        \item If $\sR^l=\sR^r=0$ and $\sT\neq 1$, $v(x)$ is the sum of four explicitly constructible unidirectionally invisible finite-range potentials with mutually disjoint supports.
        \item If $|\sR^l|+|\sR^r|\neq 0$, $v(x)$ is the sum of at most three explicitly constructible unidirectionally invisible finite-range potentials with mutually disjoint supports.
        \item The support of $v(x)$ can be chosen to be on the left or the right of any point on the real line.
        \end{enumerate}
This theorem simplifies the application of the local inverse scattering prescription we outline in [$\star$]. It also underlines the central role unidirectionally invisible potentials play in scattering theory.

\ed